\documentclass[a4paper,twocolumn,pra,english,reprint, longbibliography, superscriptaddress, showkeys, showpacs=false, nofootinbib, hyphens]{revtex4}
\usepackage{color,footnote}
\usepackage{babel}
\usepackage{amsmath,amssymb}
\usepackage{mathtools,bbm}
\usepackage{subfigure}
\usepackage{graphicx}
\usepackage{svg}
\usepackage{physics,braket}
\usepackage{comment}
\usepackage{mathtools,mathrsfs}
\usepackage[normalem]{ulem}
\usepackage[varg]{txfonts} 
\usepackage[unicode=true,pdfusetitle,bookmarks=true,bookmarksnumbered=false,bookmarksopen=false,bookmarksdepth=3,breaklinks=true,pdfborder={0 0 0},backref=false,colorlinks=true]
 {hyperref}
\usepackage{nameref} 
\usepackage{cleveref}
\usepackage[normalem]{ulem}
\makeatletter
\@ifundefined{textcolor}{}
{%
 \definecolor{BLACK}{gray}{0}
 \definecolor{WHITE}{gray}{1}
 \definecolor{RED}{rgb}{1,0,0}
 \definecolor{GREEN}{rgb}{0,1,0}
 \definecolor{BLUE}{rgb}{0,0,1}
 \definecolor{CYAN}{cmyk}{1,0,0,0}
 \definecolor{MAGENTA}{cmyk}{0,1,0,0}
 \definecolor{YELLOW}{cmyk}{0,0,1,0}
}
\pdfoutput=1
\hypersetup{colorlinks=true,citecolor=blue,linkcolor=cyan,urlcolor=blue,filecolor= green, breaklinks=true}
\usepackage{url}

\begin{document}
\author{J. S. Ara\'ujo\texorpdfstring{\href{https://orcid.org/0000-0002-6086-7189}{\includegraphics[scale=0.05]{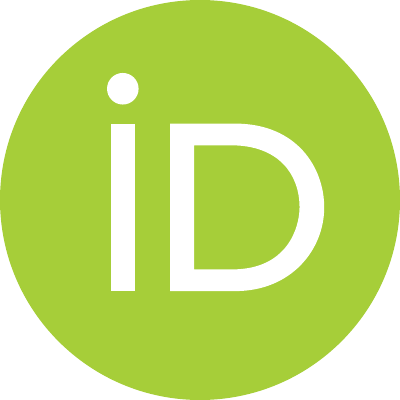}}}{ (ORCID: 0000-0002-6086-7189)}}
\email{jailson.araujo52.js@gmail.com}
\affiliation{Instituto de F\'isica, Universidade Federal do Rio de Janeiro,
Caixa Postal 68528, Rio de Janeiro, RJ 21941-972, Brazil}

\author{K. Khan\texorpdfstring{\href{https://orcid.org/0000-0001-7914-5488}{\includegraphics[scale=0.05]{orcidid.pdf}}}{ (ORCID: 0000-0001-7914-5488)}}
\affiliation{Instituto de F\'isica, Universidade Federal do Rio de Janeiro,
Caixa Postal 68528, Rio de Janeiro, RJ 21941-972, Brazil}


\author{A. S. Coelho\href{https://orcid.org/0000-0002-1860-6743}{\includegraphics[scale=0.05]{orcidid.pdf}}}
\affiliation{Instituto de F\'isica, Universidade Federal do Rio de Janeiro,
Caixa Postal 68528, Rio de Janeiro, RJ 21941-972, Brazil}
\affiliation{Department of Mechanical Engineering, Federal University of Piauí, Teresina 64049-55, Piauí, Brazil}

\selectlanguage{english}

\title{Programmable interferometer: an application in quantum channels}

\begin{abstract}
Quantum optics plays a crucial role in developing quantum computers on different platforms. In photonics, precise control over light’s degrees of freedom—including discrete variables (polarization, photon number, orbital angular momentum) and continuous variables (phase, amplitude quadratures, frequency)—is fundamental. Our model manipulates photonic systems to encode and process quantum information via the photon’s spatial degree of freedom, employing polarization as an auxiliary qubit. We propose a programmable photonic circuit that simulates quantum channels, including phase-damping, amplitude-damping, and bit-flip channels, through adjustable interferometric parameters. Furthermore, the interferometer extends to more complex channels, such as the squeezed generalized amplitude damping. This work contributes to advancing quantum simulation techniques and serves as a foundation for exploring quantum computing applications, while highlighting pathways for their practical implementation.

\end{abstract}

\date{\today}

\maketitle

\section{Introduction}

Research on quantum systems has expanded significantly, driven mostly by the second-generation of quantum technologies. Yet, complete isolation from environmental interactions remains unattainable  \cite{zurek1991decoherence, ZurekModernRev, Rivas12, Breuer02}, leading to decoherence and noise. These phenomena compromise the fidelity of quantum operations, hindering the development of scalable quantum computers \cite{Fowler12, Kitaev03} and sensors \cite{Escher11}. Understanding decoherence mechanisms and system-environment interactions enables strategies to protect quantum states, enhancing device efficiency and robustness \cite{miao2020universal,ramsay2023coherence}. Quantum error correction is a key strategy to mitigate decoherence. It encodes information to enable error detection and correction without direct state measurement, preventing wavefunction collapse \cite{Shor95, Steane06}. Decoherence-free subspaces offer another solution for well-characterized noise \cite{Lidar03}. Complementary approaches include dynamical decoupling  \cite{viola09}, which suppresses noise via timed control pulses, and environmental engineering \cite{ZurekModernRev}, which tailors system-environment interactions. In the former, precisely timed pulses reduce the effects of environmental noise on a quantum system. The latter designs the environment to minimize interactions that lead to decoherence. Despite theoretical robustness, practical implementation demands rigorous validation. Versatile platforms enable systematic study of noise and decoherence, critical for identifying discrepancies between theoretical predictions and experimental realizations \cite{Asalles2008, khan2023quantum, khan2022coherent, Araujo24, Aguilar14a}. Such platforms span multiple architectures, including trapped ions, atoms, superconducting circuits, and spins in semiconductor. Among these, photonics stands out for simulating qubit channels\cite{Wang13, Lu17, MCCUTCHEON18}. 

 Here, we present a programmable linear-optical circuit capable of simulating quantum channels—including phase-damping, amplitude-damping, and bit-flip channels—through adjustable interferometric parameters \cite{LinearOpticsComp}. This approach achieves precise control over system and environment variables via the Kraus formalism \cite{chuang1997}, enabling flexible qubit manipulation. By encoding qubits in photon paths and using polarization as an auxiliary degree of freedom (DoF), our platform offers attractive configurability, as demonstrated by its implementation of diverse quantum channels. Experimental validation of the generalized amplitude damping channel (GAD) highlights the robustness and feasibility of our all-optical framework  \cite{khan2022coherent}. This work provides theoretical and experimental insights in photonic quantum simulation, pointing out its potential to contribute to the development of quantum technologies.
 
Our paper is structured as follows. Sec.~II permits understanding the configuration of a programmable interferometer, starting with a focus on implementing the initial product state. This establishes a foundation for understanding the complex dynamics of open quantum systems using the Kraus formalism. In addition, we demonstrate our methodology through quantum state tomography via path encoding. This approach comprehensively measures quantum states by performing projections onto coherent bases. Section III extends our investigation to implementing different quantum channels.  Section IV shifts the focus toward the GAD, where we elaborate on experimental implementations using photonic systems, aiming to demonstrate the feasibility of our optical circuit design in simulating quantum communication channels. Lastly, we finalize in Sec.~V with the conclusion of the work.
\section{Programmable Interferometer}
\label{ProgrammableInterferometer}
\subsection{The Product State}
\label{P_I_product_state}
To study open quantum systems, we employ the Kraus operator formalism, which models non-unitary dynamics. Starting from a separable system-environment state
\begin{equation}
\hat{\rho}_{SE} = \hat{\rho}_S \otimes \hat{\rho}_E, \label{produto_se}
\end{equation}
where \(\hat{\rho}_S\) and \(\hat{\rho}_E = \exp(- \hat{H}_E/k_BT)/Z\) denote the density matrices of the system and the environment, respectively. Here,  \(k_B\) is the Boltzmann constant, \(T\) is the temperature and \(Z\) represents the partition function, thus describing the thermal state of the environment.
It is worth noting that while \(\hat{\rho}_E\) is typically a mixed state, it can also be a pure state in certain scenarios, depending on the preparation of the initial state. The total Hilbert space is \(\mathcal{H} = \mathcal{H}_s \otimes \mathcal{H}_e \otimes \mathcal{H}_p,\) where \( \mathcal{H}_s \), \( \mathcal{H}_e \) and \( \mathcal{H}_p \) correspond to the system, environment, and auxiliary qubit subspaces, respectively. After defining the combined space, we perform a partial trace over the auxiliary subspace \( \mathcal{H}_p \). This mathematical step removes the auxiliary DoF, ensuring the resulting state remains consistent and properly confined to the physical Hilbert space of the system and environment \(\mathcal{H}_s \otimes \mathcal{H}_e\).

\begin{figure}[htbp]
\centering 
\includegraphics[width=0.90\linewidth]{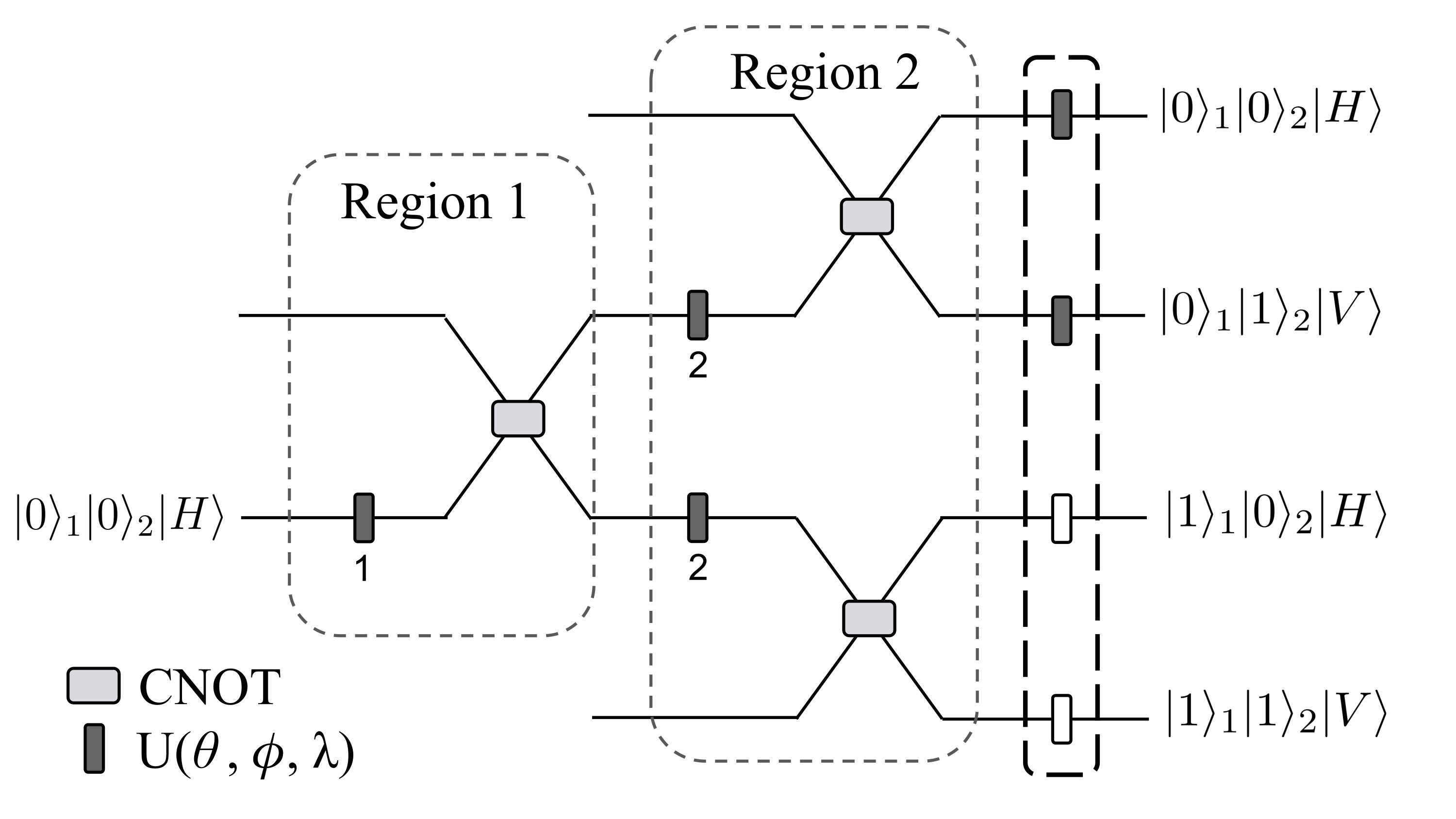}
\caption{Product state implementation by CNOT gates in conjunction with U3 gates, denoted as \(U(\theta,\phi,\lambda)\). In the circuit, Unitary 1 controls \(\alpha_1\) and \(\beta_1\), which are the probability amplitudes related to the system. Unitaries \(U_2(\theta, 0, \pi)\) control \(\alpha_2\) and \(\beta_2\), which are the probability amplitudes related to the environment. 
The unitaries encircled with the dashed line adjust the polarization. The gray unitaries apply \(U(0,0,\pi)\), while the white unitaries apply \(U(\pi,0,\pi)\). }
\label{Fig:estado_produto}
\end{figure}

Photonic systems encode quantum states via photon polarization and path DoF. Here, the system-environment product state in Eq (\ref{produto_se}) is encoded in path modes \(\ket{0}_i\) and \(\ket{1}_i\) (upper/lower paths in region \(i=1,2\) of Fig. \ref{Fig:estado_produto}), while the horizontal \((\ket{H}\)) and vertical (\(\ket{V}\)) polarization serve as an auxiliary qubit. If the photon is vertically polarized, its path changes, whereas if it is horizontally polarized, the path remains unchanged.

The circuit is initialized with a single photon in \(\ket{\psi_0}=\ket{H}\ket{0}_1\ket{0}_2\) (see Fig. \ref{Fig:estado_produto}).
Unitary transformations manipulate the polarization to populate distinct path states, to transform the initial state into a specific state on the surface of the Bloch sphere. Each transformation is represented by a unitary matrix $U_l$, corresponding to the $l$-th interaction, defined as:
\begin{eqnarray}
U_l(\theta_l,\phi_l,\lambda_l) &=&
\begin{pmatrix}
\cos(\theta_l/2) & -e^{i\lambda_l}\sin(\theta_l/2)\\
e^{i\phi_l}\sin(\theta_l/2) & e^{i(\lambda_l+\phi_l)}\cos(\theta_l/2)
\end{pmatrix} \label{matrizu1}\\&=&\begin{pmatrix}
\alpha_l & -e^{i\lambda_l}\beta_l\\
e^{i\phi_l}\beta_l & e^{i(\lambda_l+\phi_l)}\alpha_l
\end{pmatrix}
\end{eqnarray}
where $\alpha_l=\cos(\theta_l/2)$, $\beta_l=\sin(\theta_l/2)$ and \(|\alpha_l|^2+|\beta_l|^2=1\). The parameters \(\theta_l\), \(\phi_l\), and \(\lambda_l\) control the amplitude and phase shifts, corresponding to the standard U3 gate. Applying \(U_1(\theta_1,0,\pi)\) transforms the initial state \(\ket{\psi_0}\) in to
\begin{equation}
\ket{\psi_1} = (\alpha_1 \ket{H} + \beta_1 \ket{V}) \ket{00}_{12}. \label{qbit0}
\end{equation}
To populate different optical paths, we use a Controlled-NOT (CNOT) gate, which enables the generation of entanglement between the polarization and path. A polarizing beam splitter (PBS) effectively achieves this entanglement, resulting in
\begin{equation}
\ket{\psi_1'} = (\alpha_1 \ket{H}\ket{0}_1 + \beta_1 \ket{V}\ket{1}_1)\ket{0}_2. \label{cod_psi}
\end{equation}
Here, the system is codified in the path DoF $\{\ket{0}_1, \ket{1}_1\}$. Further encoding into the environmental state involves applying \(U_2(\theta,0,\pi)\) along the paths \(\ket{0}_1\) and \(\ket{1}_1\), respectively, each followed by a CNOT gate. This results in
\begin{align}
\ket{\psi_2} &= (\alpha_1\alpha_2 \ket{00}_{12}+\beta_1 \beta_2 \ket{11}_{12})\ket{H} \nonumber \\
&\quad + (\alpha_1\beta_2\ket{01}_{12}- \beta_1\alpha_2\ket{10}_{12}) \ket{V}, 
\end{align}
with \(\{\ket{0_2}, \ket{1_2}\}\) encoding the environmental states, by construction. At the end, we apply a set of transformations in each path to adjust the polarization and phases to get the state
\begin{align}
\ket{\psi_2} &= (\alpha_1\alpha_2 \ket{00}_{12}+\beta_1 \alpha_2 \ket{10}_{12})\ket{H} \nonumber \\
&\quad + (\alpha_1\beta_2\ket{01}_{12}+\beta_1\beta_2\ket{11}_{12}) \ket{V}. \label{cod_psi2.1}
\end{align}
Figure \ref{Fig:estado_produto} shows the optical circuit that generates the state described in Eq.(\ref{cod_psi2.1}). Tracing out polarization in Eq.(\ref{cod_psi2.1}) yields
\begin{equation}
\hat{\rho}_{SE} = 
\begin{pmatrix}
|\alpha_1|^2 & \alpha_1^*\beta_1 \\
\alpha_1\beta_1^* & |\beta_1|^2
\end{pmatrix}
\otimes
\begin{pmatrix}
|\alpha_2|^2 & 0 \\
0 & |\beta_2|^2
\end{pmatrix}, \label{matrix_se}
\end{equation}
consistent with the separable state in Eq.(\ref{produto_se}). By comparing Eq.(\ref{matrix_se}) with Eq.(\ref{produto_se}), we obtain \(|\alpha_1|^2 = \exp(- E_1/k_BT)/Z\) and \(|\beta_1|^2 = \exp(- E_2/k_BT)/Z\), where $E_1$ and $E_2$ are eigen-energies. This result shows that the parameters \(\alpha_1\) and \(\beta_1\) control the ambient temperature.
Although we have encoded the product state of a single qubit (system) with one qubit (environment) up to this point, we can extend this approach to describe the product state of \( n \) qubits (\( \hat{\rho}^{\otimes n}\)) by simply increasing the number of interactions that alter the photon's path.
With the initial system-environment product state encoded in the photonic framework, we now examine its dynamical evolution under the Kraus operator formalism.

\subsection{Evolution of the System by Kraus Formalism}
Using the photonic framework from Sec. II.A, we analyze the composite system  (\(\hat{\rho}_S \otimes \hat{\rho}_E\)) via Kraus formalism \cite{chuang1997}. The separable state \(\hat{\rho}_{SE}\) (Eq. \ref{produto_se}) evolves under a global unitary operation, given by

\begin{equation}
\rho_{SE}' = \mathcal{\hat{U}}\rho_{SE} \mathcal{\hat{U}}^\dagger, \label{eq.evolution}
\end{equation}
where \(\mathcal{\hat{U}}\) represents the global unitary operator. 
The system’s evolution is obtained by tracing out the environment, leading to the Kraus representation
\begin{equation}
\Lambda(\rho_S) = \sum_{\mu} M_{\mu}\rho_S M_{\mu}^\dagger, \label{super_evolution}
\end{equation}
where \(M_{\mu}= \bra{\mu}\mathcal{\hat{U}}\ket{0}\) denotes the Kraus operators, with \(\{\ket{\mu}\}\) constituting an orthonormal basis for the environment. These operators, acting on the system's state \(\rho_S\), are derived from projecting the unitary evolution onto an orthonormal basis of the environment. 

The Kraus operators capture the effects of all possible environmental interactions, with their sum ensuring total probability preservation, \(\text{Tr}(\Lambda(\rho_S)) = 1\), and satisfying the completeness relation \(\sum_{\mu} M_{\mu}^\dagger M_{\mu} = I\). We obtain different sets of equivalent Kraus operators in other bases by performing the partial trace operation on Eq.(\ref{eq.evolution}). 

\begin{figure}[htbp]
\centering 
\includegraphics[width=0.99\linewidth]{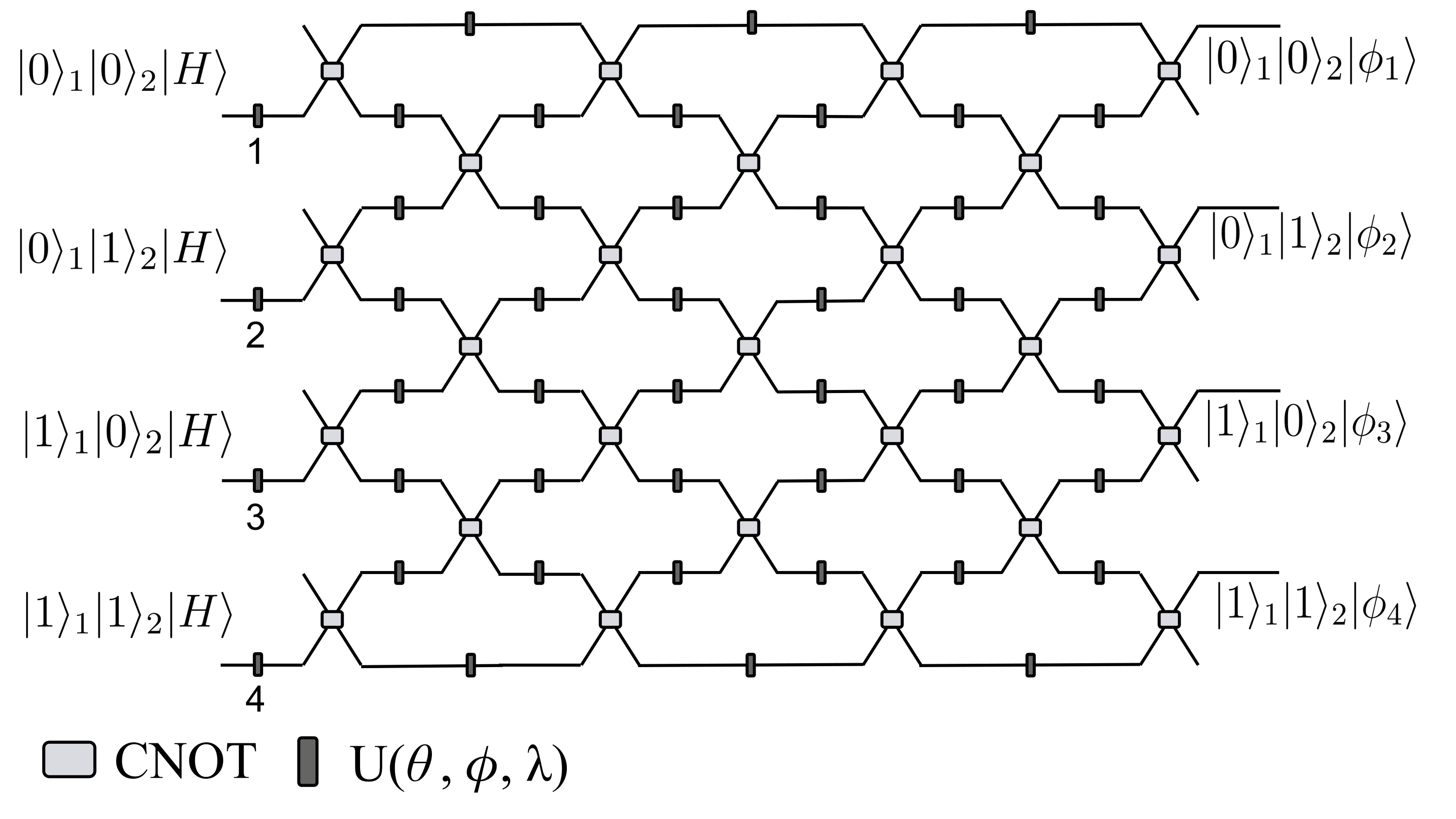}
\caption{Schematic illustration of an optical circuit featuring an arrangement of CNOT gates and local unitary gates \(U(\theta, \phi, \lambda)\). Unitary gates 1 to 4 include a free parameter \(\theta\), which influences the probabilities of the resulting states, $\phi$ and $\lambda$ will generally remain constant. Other unitary gates typically assume fixed binary values of \(U(0,0,\pi)\) or \(U(\pi,0,\pi)\) degrees. This configuration allows for the control of the transformation probabilities of the input states, enabling various quantum operations.}
\label{Fig:canal}
\end{figure}

Every quantum operation that acts on a system in a Hilbert space of $d$ dimensions can be characterized by $d^2$ Kraus operators \cite{chuang1997}. This property leads us to the following unitary map, which describes the evolution of the system \cite{Asalles2008}
\begin{eqnarray}
\ket{\phi_0}\ket{0} &\rightarrow& M_0\ket{\phi_0}\ket{0}+\cdots+ M_{d^2-1}\ket{\phi_0}\ket{d^2-1};\nonumber\\
\ket{\phi_1}\ket{0} &\rightarrow& M_0\ket{\phi_1}\ket{0}+ \cdots+ M_{d^2-1}\ket{\phi_1}\ket{d^2-1};\nonumber\\
\vdots\nonumber\\
\ket{\phi_{d-1}}\ket{0} &\rightarrow& M_0\ket{\phi_{d-1}}\ket{0}+\cdots+ M_{d^2-1}\ket{\phi_{d-1}}\ket{d^2-1},\nonumber \\ \label{map}
\end{eqnarray}
where each \(M_i\) operates within the system's subspace. The general map in Eq. (\ref{map}) can be implemented by the interferometer in Fig. \ref{Fig:canal}.  To illustrate the practical application of these concepts, consider the interaction between a qubit and a thermal qubit, as per Eq.(\ref{matrix_se}), we have
\begin{align}
\sqrt{\alpha_2}\ket{0}\ket{0} &\rightarrow M_{00}\ket{0}\ket{0} + M_{01}\ket{0}\ket{1},\nonumber\\
\sqrt{\alpha_2}\ket{1}\ket{0} &\rightarrow M_{00}\ket{1}\ket{0} + M_{01}\ket{1}\ket{1},\nonumber\\
\sqrt{\beta_2}\ket{0}\ket{1} &\rightarrow M_{10}\ket{0}\ket{0} + M_{11}\ket{0}\ket{1},\nonumber\\
\sqrt{\beta_2}\ket{1}\ket{1} &\rightarrow M_{10}\ket{1}\ket{0} + M_{11}\ket{1}\ket{1},  \label{map_qbit}
\end{align}
where \(M_{ij} = \sqrt{\gamma_j}\langle i | \mathcal{U} | j \rangle \) and \(\gamma_j\) is defined such that
\begin{equation}
\gamma_j = \begin{cases} \alpha_2 & \text{if } j = 0 \\ \beta_2 & \text{if } j = 1 \end{cases}.    
\end{equation}
Once the Kraus operators \( M_{ij} \) are specified, we determine the transitions between modes, allowing us to implement the map in Eq. (\ref{map}) using the optical circuit shown in Figure \ref{Fig:canal}, which encompasses all possible transitions between modes. The design of this circuit is not unique; it is necessary only to maintain the optical path length identical for each mode and ensure the possibility of path switching among all modes.
A sequence of interferometers can implement this scheme by coherently separating and recombining light, which ensures coherent superposition between modes. The circuit shown in Figure \ref{Fig:canal} includes a set of CNOT gates and unitary operations \(U(\theta,\phi,\lambda)\). This combination of gates facilitates transitions between the modes specified by the map in Eq.(\ref{map_qbit}). 

Ultimately, the output modes exhibit different contributions from previous modes, meaning the polarization can be represented as a combination of polarizations \(\ket{H}\) and \(\ket{V}\), denoted in the circuit as \(\ket{\phi_i}\). The polarization of the circuit's output modes must match that of the input modes encoding the initial state, as indicated in Eq. (\ref{cod_psi2.1}). For example, if the initial mode is \(\ket{01H}\), then the output modes originating from this input mode must retain the same polarization, i.e., \(\ket{01H} \rightarrow \sqrt{1-p}\ket{01H} + \sqrt{p}\ket{10H}\). This criterion ensures the consistency of the maps presented.

Since the circuit design does not have a unique configuration, it allows for the implementation of more general dynamics. Using the method presented here, we can construct a circuit capable of realizing a broader class of dynamics. In the simple case of a system with four path levels, we keep the input and output fixed to four modes, but we can increase the number of internal paths within the circuit. This expansion would enable more complex dynamics, leading to more sophisticated mixtures of the input modes.

Having established the dynamics of the system under the Kraus formalism, we now turn to the experimental characterization of these states. State tomography in path-encoded qubits enables the reconstruction of the density matrix, providing a direct measure of the system’s quantum state.

\subsection{State Tomography in Path Encoding qubit}
\label{State_Tomography}

\begin{figure}[htbp]
\centering 
\includegraphics[width=0.99\linewidth]{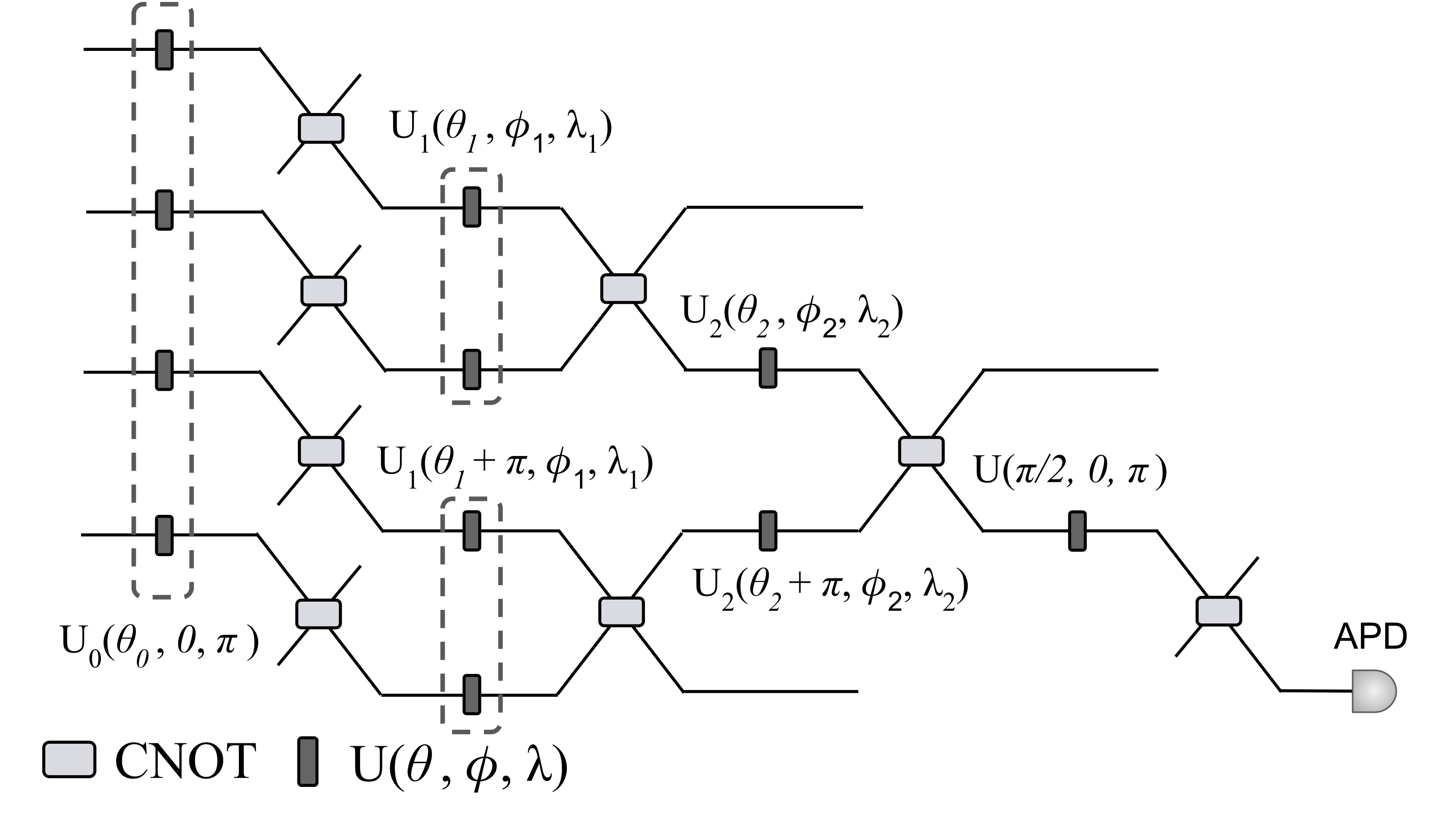}
\caption{Illustration of the projection circuit: Projections are performed onto the path degrees of freedom, with each unitary controlled individually to allow coherent combinations of the paths. The initial plates (encircled by the dashed line) and the subsequent CNOT gate are adjusted to select either H or V polarizations, which leaves the polarization separable. Thus, partial tracing over this DoF is direct. Notably, this configuration employs only one Avalanche Photodiode Detector (APD).}
\label{Fig:projecao}
\end{figure}

Two-qubit state tomography requires 16 projective measurements \cite{james2001measurement}. When the qubits are encoded in the polarization, a set of wave plates and polarizers are adjusted to specific angles following the protocol in Ref. \cite{james2001measurement}. For codification in the path, we map the path onto polarization using beam desplacers or polarized-beam-splitters and then implement the same protocol above. Figure \ref{Fig:projecao} depicts the optical circuit for path-encoded tomography. Polarization control directs each mode to the detector or combines multiple modes coherently. While similar setups have been used \cite{Farias12, Aguilar14b, Aguilar14a, Araujo24}, our design deffers by tailoring to the specific requirements of path-encoded qubits.

The circuit projects onto \(\ket{H}\) and \(\ket{V}\) polarizations, enabling a partial trace over polarization by summing detection counts. To avoid distinguishability from the path qubit’s polarization tag, we project the final state onto a diagonal basis before detection. The circuit’s projections are described by \cite{james2001measurement}:
\begin{eqnarray}
    \ket{\psi(\theta_1,\phi_1,\lambda_1,\theta_2,\phi_2,\lambda_2)}=U_1(\theta_1,\phi_1,\lambda_1)\otimes\nonumber \\U_2(\theta_2,\phi_2,\lambda_2)\cdot \ket{00}.\label{eq.pro}
\end{eqnarray}
where \(U_1\) and \(U_2\) act on qubits 1 and 2, respectively. By adjusting the parameters $\theta$, $\phi$, and $\lambda$ we configure the unitaries to project onto specific states for detection.
The table \ref{tabela} lists the 16 configurations of \(U_1\) and \(U_2\) required for full state tomography. With the state tomography method for path-encoded qubits established, we next demonstrate its utility in simulating specific quantum channels, beginning with the dephasing channel.
\begin{table}
\caption{The tomographic analysis. The projection measurements provide a set of 16 data points that allow the density matrix of the two-qubit state to be reconstructed. We use the notation \( |D\rangle =  ({|H\rangle + |V\rangle})/{\sqrt{2}}  \) and \( |R\rangle = ({|H\rangle + i|V\rangle})/{\sqrt{2}} \).}
\begin{center}
\begin{tabular}{|c|c|c|c|c|c|c|c|c|c|c|c|c|c|c|c|}
\hline
n & qubit$_1$ & qubit$_2$  & $U_1(\theta_1,\phi_1,\lambda_1)$ & $U_2(\theta_2,\phi_2,\lambda_2)$\\
\hline
1 & H & H & $U_1(0,0,\pi)$ & $U_2(0,0,\pi)$\\
2 & H & V &  $U_1(0,0,\pi)$ & $U_2(\pi,0,\pi)$\\
3 & V & V &  $U_1(\pi,0,\pi)$ & $U_2(\pi,0,\pi)$\\
4 & V & H &  $U_1(\pi,0,\pi)$ & $U_2(0,0,\pi)$\\
5 & R & H &   $U_1(\pi/2,\pi/2,\pi/2)$ & $U_2(0,0,\pi)$\\
6 & R & V &   $U_1(\pi/2,\pi/2,\pi/2)$ & $U_2(\pi,0,\pi)$\\
7 & D & V &  $U_1(\pi/2,0,\pi/2)$ & $U_2(\pi,0,\pi)$\\
8 & D & H &  $U_1(\pi/2,0,\pi/2)$ & $U_2(0,0,\pi)$\\
9 & D & R &   $U_1(\pi/2,0,\pi/2)$ & $U_2(\pi/2,\pi/2,\pi/2)$\\
10 & D & D & $U_1(\pi/2,0,\pi/2)$ & $U_2(\pi/2,0,\pi/2)$\\
11 & R & D & $U_1(\pi/2,\pi/2,\pi/2)$ & $U_2(\pi/2,0,\pi/2)$\\
12 & H & D & $U_1(0,0,\pi)$ & $U_2(\pi/2,0,\pi/2)$\\
13 & V & D & $U_1(\pi,0,\pi)$ & $U_2(\pi/2,0,\pi/2)$\\
14 & V & R &  $U_1(\pi,0,\pi)$ & $U_2(\pi/2,\pi/2,\pi/2)$\\
15 & H & R & $U_1(0,0,\pi)$ & $U_2(\pi/2,\pi/2,\pi/2)$\\
16 & R & R & $U_1(\pi/2,\pi/2,\pi/2)$ & $U_2(\pi/2,\pi/2,\pi/2)$\\
\hline
\end{tabular} 
\label{tabela}
\end{center}
\end{table}
\section{Application in Quantum Channels}

\subsection{Dephasing Channels}
The dephasing channel models energy-preserving interactions between a quantum system and its environment, where quantum coherence decays without energy exchange. This process captures the loss of phase information while maintaining the population of quantum states. The dephasing channel is usually defined by the Kraus operators
\begin{equation}
M_0 = 
\begin{pmatrix}
1 & 0 \\
0 & \sqrt{1-p}
\end{pmatrix},
\quad
M_1 = 
\begin{pmatrix}
0 & 0 \\
0 & \sqrt{p}
\end{pmatrix},
\end{equation}
where $p \in [0,1]$  represents the decoherence probability. The mapping concisely that captures the transformation effected by these operators is given by
\begin{align}
\ket{00} &\rightarrow \ket{00}, \\
\ket{10} &\rightarrow \sqrt{1-p}\ket{10} + \sqrt{p}\ket{11},
\end{align}
indicating that the state $\ket{00}$ remains unaffected, whereas the state $\ket{10}$ undergoes a probabilistic transition, retaining its original form with probability $1-p$ or transitioning to $\ket{11}$ with probability $p$.

To implement the dephasing channel, we configure the photonic circuit(Fig. \ref{Fig:canal}) by setting identical optical path lengths and activating specific unitary operations. This ensures coherent transitions between modes, as required by the Kraus operators \(M_0\) and \(M_1\). In this context, under the assumption of zero temperature (implying $\beta_2 = 0$), the system-environment density matrix Eq.(\ref{matrix_se}) simplifies to
\begin{equation}
\rho_{SE} = 
\begin{pmatrix}
|\alpha_1|^2 & \alpha_1^{*}\beta_1 \\
\alpha_1\beta_1^{*} & |\beta_1|^2
\end{pmatrix}
\otimes
\begin{pmatrix}
1 & 0 \\
0 & 0
\end{pmatrix}. \label{dephase_se}
\end{equation}
This product state is encodable through the precise adjustment of unitaries within the circuit, as shown in Fig. \ref{Fig:estado_produto}, specifically setting unitaries 2 and 3 to $U(0,0,\pi)$ and $U(\pi,0,\pi)$, respectively, with unitary 1 governed by the adjustable parameter $\theta$. To effectuate the transformation illustrated in the map, it becomes necessary to determine the configuration of the 42 unitary operations in the circuit (Fig. \ref{Fig:canal}). Although this task may seem complex, in practice, the requirement is limited to finding the configuration of the unitary operations specific to the modes under analysis. This approach simplifies the process by focusing on the elements directly relevant to propagation.

\begin{figure}[htbp]
\centering 
\includegraphics[width=0.90\linewidth]{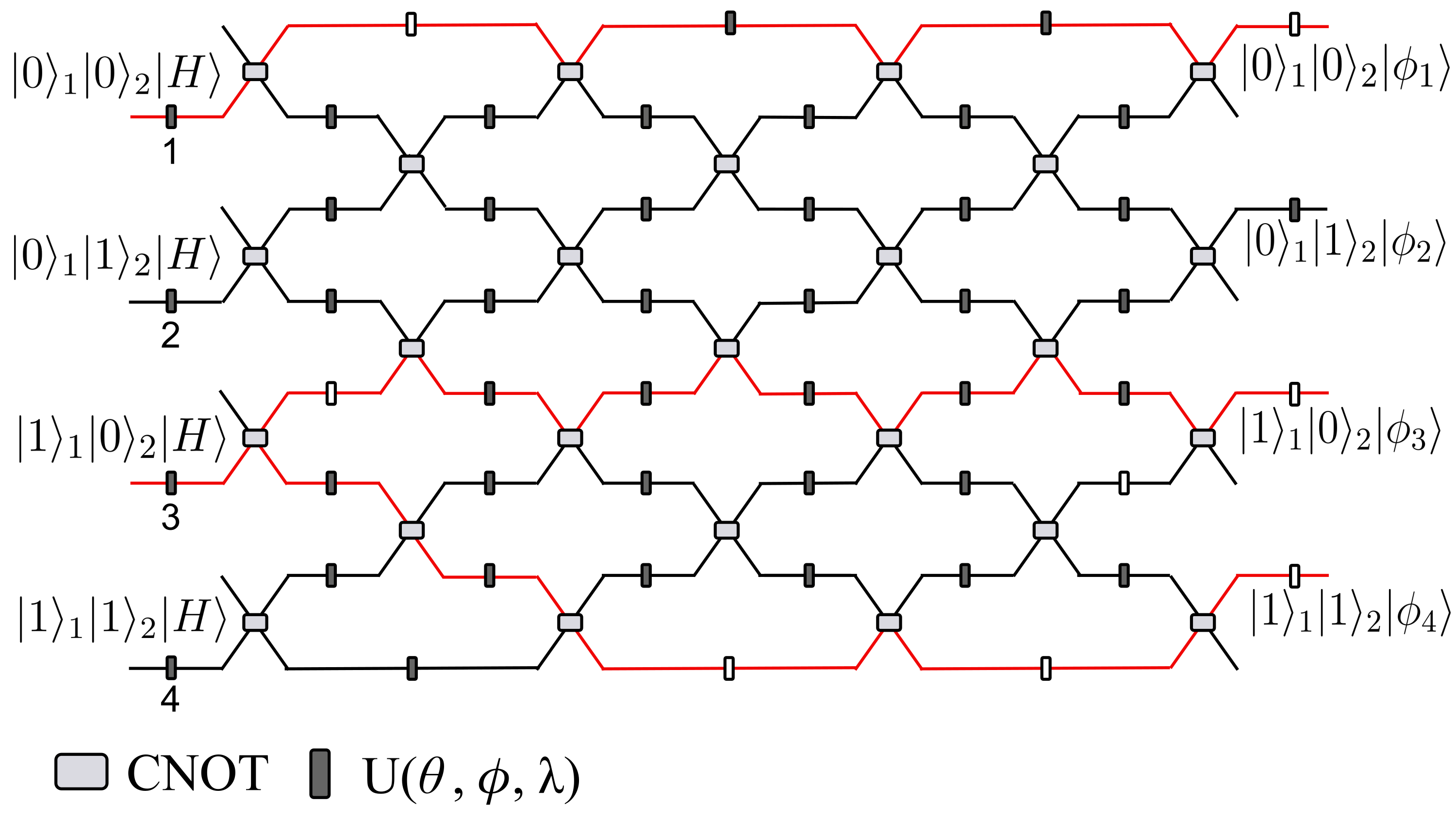}
\caption{The circuit is adjusted to implement the dynamics of the phase-damping channel. 
In this setup, the unitary operation in white alters the mode guide, causing the transition from $\ket{0}(\ket{1})$ to $\ket{1}(\ket{0})$, while the unitary operation in black leaves the mode guide unchanged. Only unitaries 1 and 3 are active: unitary 3 controls the channel parameter \( p = \cos^2(\theta/2) \), and we set the unitary 1 to \( U(0, 0, \pi) \). Only a few unitaries are active throughout the circuit, specifically those along the path of the illuminated modes shown in red. In this channel, \( \ket{\phi_1} = \alpha_1\ket{H}\) , \( \ket{\phi_3} = \beta_1\sqrt{1-p}\ket{H}\) and  \( \ket{\phi_4} =\beta_1\sqrt{p}\ket{H} \) corresponds to vertical polarization.}
\label{Fig:dephase_channel}
\end{figure}

In Fig. \ref{Fig:dephase_channel}, the circuit is configured to implement a phase-damping channel, highlighting how specific configurations correspond to particular maps. 
This straightforward example vividly demonstrates the effectiveness and versatility of the proposed programmable optical circuit. The optical processor goes beyond basic operations, effectively implementing more complex channels, such as the generalized amplitude damping channel and the squeezed generalized amplitude damping channel \cite{chuang1997}.

Beyond phase coherence loss, quantum systems in thermal environments exhibit energy relaxation. The generalized amplitude damping channel provides a framework to model these dynamics.

\subsection{Generalized amplitude damping channel}
\label{GAD}
The generalized amplitude damping (GAD) channel extends the amplitude damping model to non-zero temperature environments, accounting for thermal excitations and relaxation processes. This model expands the principles of the amplitude damping channel by incorporating environmental states that are not exclusively ground states. Four Kraus operators characterize the GAD channel
\begin{eqnarray}
M_{00} =
\sqrt{\alpha_2}  
\begin{pmatrix}
1 & 0\\
0 & \sqrt{1-p}
\end{pmatrix};
\quad
M_{01} =
\sqrt{\alpha_2}  
\begin{pmatrix}
0 & \sqrt{p}\\
0 & 0
\end{pmatrix};
\\
M_{10} =
\sqrt{\beta_2}  
\begin{pmatrix}
0 & 0\\
\sqrt{p} & 0
\end{pmatrix};
\quad
M_{11} =
\sqrt{\beta_2}  
\begin{pmatrix}
\sqrt{1-p} & 0\\
0 & 1
\end{pmatrix},
\end{eqnarray}
where $\alpha_2$ ($\beta_2$) corresponds to the ground (excited) state population of the thermal bath. The map that describes this channel, according to Eq.(\ref{map_qbit}), is given by
\begin{eqnarray}
\ket{0}_s\ket{0}_E &\longrightarrow &\quad  \ket{0}_s\ket{0}_E ;\nonumber \\
\ket{1}_s\ket{0}_E &\longrightarrow & \quad \sqrt{1-p}\ket{1}_s\ket{0}_E + \sqrt{p}\ket{0}_s\ket{1}_E;\nonumber \\
\ket{0}_s\ket{1}_E &\longrightarrow & \quad \sqrt{1-p}\ket{0}_s\ket{1}_E + \sqrt{p}\ket{1}_s\ket{0}_E;\nonumber\\
\ket{1}_s\ket{1}_E &\longrightarrow &\quad  \ket{1}_s\ket{1}_E. \label{ADG4} 
\end{eqnarray}
Given that the environment can initiate the interaction in an ``excited" state, this description considers a probability $p$ of the environment exciting the system or vice versa. Moreover, no transitions will occur if the system and the environment are already excited or in the ground state.

\begin{figure}[htbp]
\centering 
\includegraphics[width=0.99\linewidth]{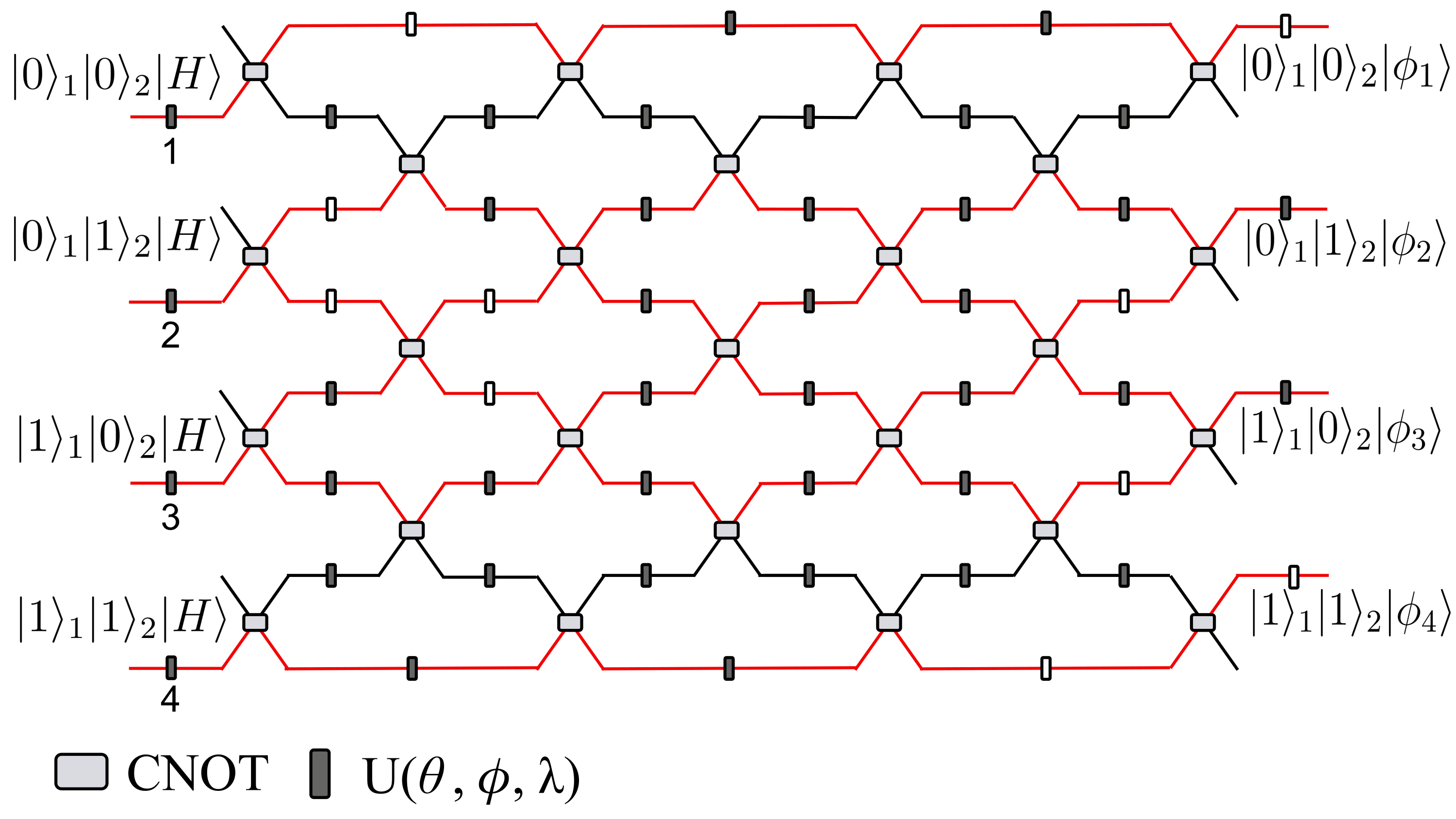}
\caption{The quantum circuit shown here implements the dynamics of the Generalized Amplitude Damping channel. The circuit includes four gates (1, 2, 3, 4) with a free parameter $\theta$. Unitaries 2 and 3 control the channel parameter \( p = \cos^2(\theta) \), while unitaries 1 and 4 are set to \( U(0,0,\pi) \) and \( U(\pi,0,\pi) \), respectively. Furthermore, along the circuit, the unitaries that lie along the illuminated modes (represented by the color red) assume two states: \( U(0,0,\pi) \) (depicted in black) and \( U(\pi,0,\pi) \) (depicted in white).
For this channel, we find that \( \ket{\phi_1} = \alpha_1\alpha_2\ket{H}\) and \( \ket{\phi_4} =\beta_1\beta_2\ket{V} \) while \( \ket{\phi_2} = \alpha_1\beta_2\sqrt{1-p}\ket{V}+\alpha_2\beta_1\sqrt{p}\ket{H} \) and \(  \ket{\phi_3} = \alpha_2\beta_1\sqrt{1-p}\ket{H}+\alpha_1\beta_2\sqrt{p}\ket{V} \).}

\label{Fig:gad_channel}
\end{figure}

Furthermore, Figure \ref{Fig:estado_produto} visually supports this comprehensive approach by illustrating the interaction between the system and the environment at a temperature above zero through unitaries labeled 1, 2, and 3. These unitaries use parameters $\theta_i$ to control the system and the environment's temperature. Figure \ref{Fig:gad_channel} shows the optical circuit configured to simulate the dynamics of the GAD. Given the complexity of the channel, simulating the evolution of the system's dynamics requires activating a more significant number of elements in the circuit.

This advanced implementation showcases the optical circuit as a versatile experimental platform for exploring complex quantum channels like the GAD. It enhances our understanding of key quantum phenomena such as dissipation and underscores the circuit's significant potential for advancing quantum computing technologies and foundational quantum mechanics research.

The generalized amplitude damping channel accounts for thermal effects, yet realistic environments often involve non-classical states of the reservoir. To address this, we introduce the squeezed generalized amplitude damping (SGAD) channel, which incorporates squeezing operations to engineer the reservoir’s quantum noise and control decoherence dynamics.
\subsection{Squeezed generalized amplitude damping channel}
\label{SGAD}
The squeezed generalized amplitude damping (SGAD) channel incorporates squeezing operations to modulate decay rates in quantum transitions. This allows control over the system’s relaxation dynamics, suppressing or enhancing decoherence depending on the squeezing parameters.
Considering a two-level system, with the bath initially in a squeezed thermal state, the resulting dynamics, governed by a Lindblad-type evolution, leads to a completely positive map that extends the concept of a generalized amplitude damping channel \cite{srikanth2008squeezed}. The SGAD channel is defined by the Kraus operators:
\begin{eqnarray}
 M_{00} &\equiv& \sqrt{\alpha_2} \begin{pmatrix} \sqrt{1 - \alpha} & 0 \\ 0 & \sqrt{1 - \beta} \end{pmatrix} ; \label{1}\\
 M_{01} &\equiv& \sqrt{\alpha_2} \begin{pmatrix} 0 & \sqrt{\beta} \\ \sqrt{\alpha}e^{-i\phi} & 0 \end{pmatrix} ; \\
 M_{11} &\equiv& \sqrt{\beta_2} \begin{pmatrix} \sqrt{1 - \mu} & 0 \\ 0 & \sqrt{1 - \nu} \end{pmatrix} ;\\
 M_{10} &\equiv& \sqrt{\beta_2} \begin{pmatrix} 0 & \sqrt{\nu} \\ \sqrt{\mu}e^{-i\lambda} & 0 \end{pmatrix}\label{4} ,   
\end{eqnarray}
where \(\alpha\), \(\beta\), \(\mu\) and \(\nu\) depend on the squeezing parameters and bath temperature. An advantage of using a squeezed thermal bath is that it can suppress the decay rate of quantum coherence, preserving nonclassical effects \cite{srikanth2008squeezed}.

\begin{figure}[htbp]
\centering 
\includegraphics[width=0.99\linewidth]{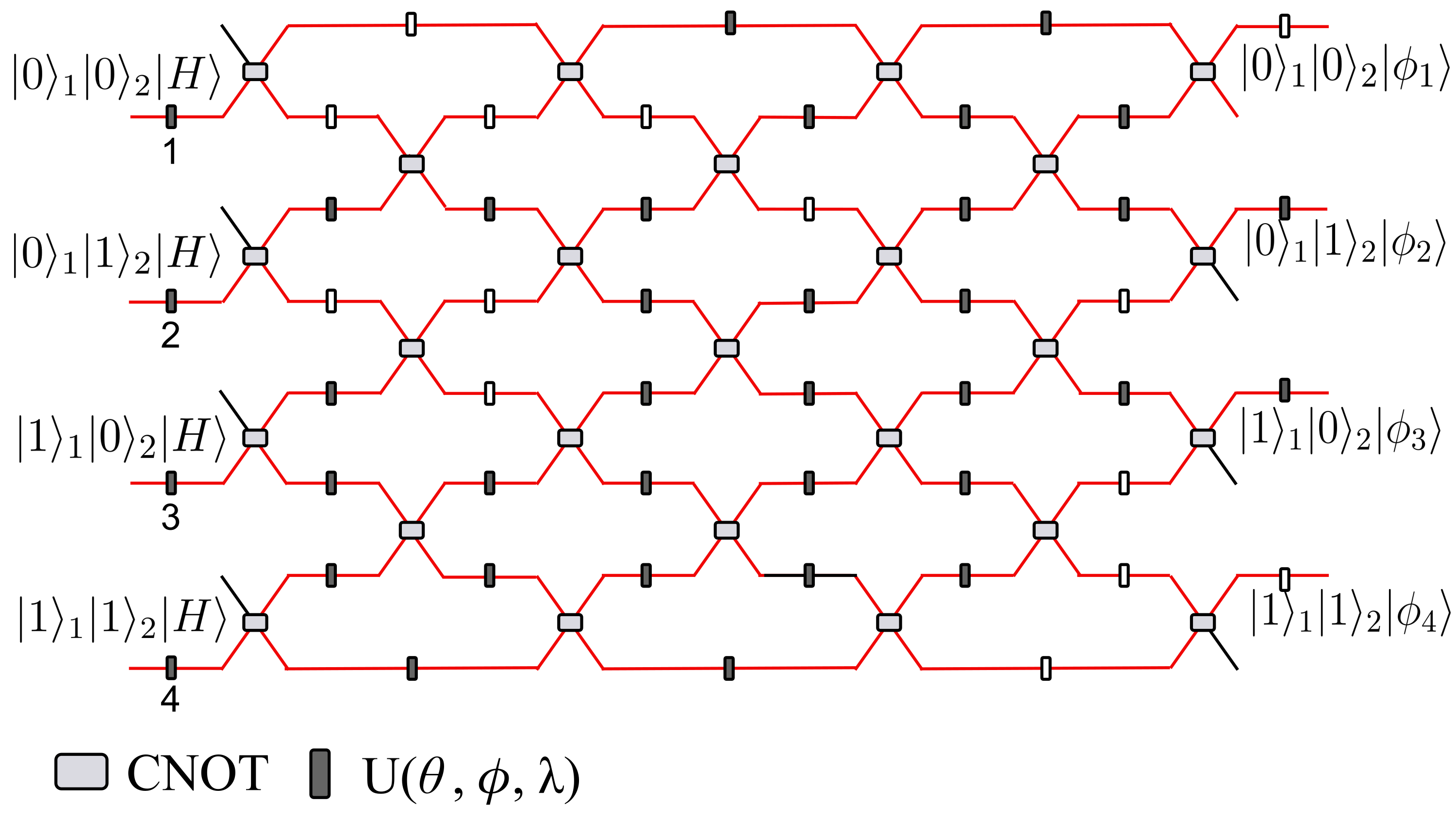}
\caption{We adjusted the circuit to implement the dynamics of the SGAD channel. In this configuration, the unitary operation represented in white is responsible for altering the mode guide, transitioning from $\ket{0}(\ket{1})$ to $\ket{1}(\ket{0})$. The unitary operation in black keeps the mode guide unchanged. Unitaries 1, 2, 3, and 4 control the transition probability, as shown in the map in Eq.(\ref{Squeezed}), where unitaries 1 and 2 are adjusted to $U(\theta_1, -\phi, 0)$ and $U(\pi+\theta_2, 0, \pi-\lambda)$, respectively, with $\theta_i$ controlling the transition probability,$\lambda$ and $\phi$ controlling the phase. Unitaries 3 and 4 are adjusted to $U(\pi+\theta_3, 0, \pi)$ and $U(\theta_4, 0, \pi)$, respectively. For this channel, we find that $\ket{\phi_1}=\alpha_1\alpha_1\sqrt{1-\alpha}\ket{H}+\beta_1\beta_2\sqrt{\nu}\ket{V}$, $\ket{\phi_2}=\alpha_1\beta_2\sqrt{1-\beta}\ket{V}+\beta_1\alpha_2\sqrt{\mu}e^{-i\lambda}\ket{H}$, $\ket{\phi_3}\sqrt{1-\mu}\ket{H}+\sqrt{\beta}\ket{V}$, and $\ket{\phi_4}=\beta_1\beta_2\sqrt{1-\nu}\ket{V}+\alpha_1\alpha_2\sqrt{\alpha}e^{-i\phi}\ket{H}$.
}
\label{Fig:sgad_channel}
\end{figure}

Applying the Kraus operators Eq.(\ref{1}-\ref{4}) to the equation (\ref{map_qbit}), we can construct the map that describes this dynamic

\begin{eqnarray}
\ket{0}_s\ket{0}_E &\longrightarrow & \sqrt{1-\alpha}\ket{0}_s\ket{0}_E+e^{-i\phi}\sqrt{\alpha}\ket{1}_s\ket{1}_E ;\nonumber \\
\ket{1}_s\ket{0}_E &\longrightarrow & \sqrt{1-\beta}\ket{1}_s\ket{0}_E + \sqrt{\beta}\ket{0}_s\ket{1}_E;\nonumber \\
\ket{0}_s\ket{1}_E &\longrightarrow & \sqrt{1-\mu}\ket{0}_s\ket{1}_E + e^{-i\lambda} \sqrt{\mu}\ket{1}_s\ket{0}_E;\nonumber\\
\ket{1}_s\ket{1}_E &\longrightarrow &  \sqrt{1-\nu}\ket{1}_s\ket{1}_E+\sqrt{\nu}\ket{0}_s\ket{0}_E. \label{Squeezed} 
\end{eqnarray}
Thus, we know all transitions between modes necessary for reproducing these dynamics using our programmable optical circuit. The SGAD exhibits characteristics quite different from the channels presented so far, such as each line of its map showing a different mode transition probability and the emergence of phase. All these parameters that appear in the map are intrinsically related to the squeezed bath, as shown in \cite{srikanth2008squeezed}, indicating that the bath squeezing is associated with the decay rate of the channel.

Figure \ref{Fig:sgad_channel} presents the configuration of the optical circuit that implements the SGAD, where the parameters follow the relations \(\alpha=\cos^2{\theta_1}\), \(\beta=\cos^2{\theta_2}\), \(\mu=\cos^2{\theta_3}\), and \(\nu=\cos^2{\theta_4}\). Although the channel has a complex theoretical formulation, its implementation in the optical circuit is relatively simple. This example illustrates the ability of the optical circuit to represent the evolution of quantum systems, including those associated with nontrivial channels.

The channels analyzed here (dephasing, GAD, SGAD) demonstrate that the platform can simulate well-defined noise processes. For more general applications, this approach also allows the simulation of Pauli channels, which involve arbitrary combinations of bit-flip, phase-flip, and depolarizing noise. These channels are an important subset of the unital channels \cite{chuang1997}. Appendix \ref{appendix} presents the details on encoding a Pauli channel in the proposed circuit.

\section{Experimental Implementation- GAD}
The objective now is to present a specific optical circuit designed for the GAD channel, highlighting its programmability aspect. By focusing on the GAD implementation, we aim to illustrate the feasibility of creating a fully programmable optical circuit using this methodology.
\begin{figure*}[htb]
\centering
\includegraphics[scale=0.3]{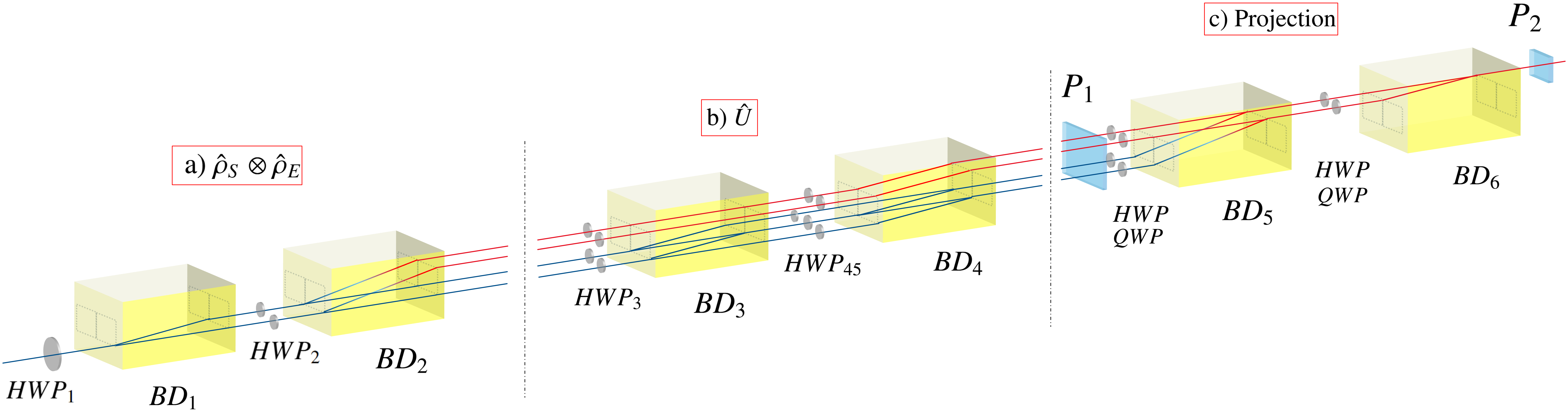}
\caption{Experimental scheme of the two-layer interferometer. a) BD1 and BD2 encode the initial state, while $HWP_1$ and $HWP_2$ control the system and environment states, respectively. b) BD3 and BD4 implement the global unitary operation that governs the transitions of the GAD channel, described by the map in Eq.(\ref{mapa_gad}). The transition parameter $p$ is adjusted by $HWP_3$. c) BD5, BD6, and the wave plates (HWP and QWP) perform projective measurements. $P_1$ consists of a half-wave plate followed by a PBS, used to select between horizontal and vertical polarization. $P_2$ projects the polarization onto $\sigma_x$.} 
\label{Fig:gad-exp}
\end{figure*}

\subsubsection{Product state}
Beam displacers (BDs), together with wave plates, implement the initial product state by spatially separating photons according to their polarization, in a manner analogous to polarizing beam splitters (PBSs). The orientation of the BD’s optical axis determines the deflection direction, allowing it to function as a CNOT gate that directs photons into distinct paths based on their polarization. As illustrated in Fig. \ref{Fig:gad-exp} a), implementing the product state requires two BDs. We insert half-wave plates (HWP) between the beam splitters (BDs) to adjust the state's polarization, an operation equivalent to that implemented by the U3 gate \(\left(U(\theta, 0, \pi)\right)\). This configuration enables the encoding of the product state, as described in Section \ref{P_I_product_state}. After BD2, the initial system-environment state corresponds to the product state specified in equation \ref{matrix_se}. 

The parameters \(\alpha_1=\cos(\theta_1)\) and \(\beta_1=\sin(\theta_1)\) represent the probability amplitudes associated with the system, while \(\alpha_2=\sin(\theta_2)\) and \(\beta_2=\cos(\theta_2)\) correspond to the probability amplitudes describing the statistical mixture of the environment. The experimental setup imposes \(\alpha_2=\sin(\theta_2)\), but this choice does not affect the analysis presented, as sine and cosine are complementary functions.  
This encoding allows full control over the system's and environment's initial state before it is subjected to the GAD channel.


\subsubsection{State evolution}


The global evolution described by the map in Eq. (\ref{ADG4}) is implemented as illustrated in Fig. \ref{Fig:gad-exp}.b, using a combination of HWPs and BDs to achieve the mode transition required by the GAD map. The transition probability between the modes \(\ket{01}\) and \(\ket{10}\) is adjusted through the half-wave plate \( \text{HWP}_3 \), with the channel parameter given by \( p = \cos^2\theta_3 \). Thus, the channel's decay rate can be controlled by varying the angle \(\theta_3\).

The utility of these optical elements in achieving the intended evolution is evident in the output of BD4. The evolved state of the system and environment, denoted as $\hat{\rho}'_{SE}$, obtained post the traversal through BD4, is represented as follows
\begin{eqnarray}
\alpha_1 \sin \theta_2 \ket{00} &\longrightarrow&     \alpha_1 \sin \theta_2 \ket{00}; \nonumber\\
\beta_1 \sin \theta_2 \ket{10} &\longrightarrow& \beta_1 \sin \theta_2 \sin \theta_3 \ket{10}+ \beta_1 \sin \theta_2 \cos \theta_3 \ket{01};\nonumber\\
\alpha_1 \cos \theta_2 \ket{01} &\longrightarrow& \alpha_1 \cos \theta_2 \sin \theta_3 \ket{01}+ \alpha_1 \cos \theta_2 \cos \theta_3 \ket{10};\nonumber\\
\beta_1 \cos \theta_2 \ket{11} &\longrightarrow&   \beta_1 \cos \theta_2 \ket{11}.  \label{mapa_gad}
\end{eqnarray}
Ultimately, the precision and strategic design of the setup shown in \ref{Fig:gad_channel} underscore its effectiveness in replicating the characteristics of the GAD. Here, one can observe that the map obtained at the output of BD4 precisely matches the required map of the GAD in  Eq. (\ref{ADG4}), showcasing the successful simulation of quantum communication channels through optical means.
\subsubsection{Projections}

To perform projective measurements on the state, as described in Section \ref{State_Tomography}, it is necessary to coherently combine different modes and select different polarizations. Fig. \ref{Fig:gad-exp}.c illustrates the implementation of this projection. 
The procedure begins with a half-wave plate (HWP) followed by a polarizing beam splitter (PBS) to select between \(H\) and \(V\) polarizations. After this selection, beam displacers (BDs), together with HWPs and QWPs, perform the projections onto the paths.  
In the final stage of the circuit, an HWP set to \(\pi/8\) and a PBS project the state onto the polarization in the \(\sigma_x\) operator, eliminating path distinguishability introduced by polarization differences. The protocol described in Table \ref{tabela} allows for the reconstruction of the global density matrix (system and environment) through state tomography.

We set the channel parameters to \(\theta_1 = \pi/8\), \(\theta_2 = \pi/8\), and \(\theta_3 = \pi/8\) and performed the tomography procedure. The obtained global density matrix (Eq.(\ref{matrixgad1})) showed approximately 95\% fidelity with the theoretically predicted matrix. This result demonstrates the consistency of the method used for the implementation of the quantum channel.
\begin{widetext}
\begin{equation}
\Lambda(\rho_{SE})=
 \begin{pmatrix}
 0.253077 & 0.178583\, -0.0174541 i & 0.129289\, -0.0237165 i & -0.0360439-0.0166259 i \\
 0.178583\, +0.0174541 i & 0.276904 & 0.225649\, -0.0334031 i & 0.15571\, -0.0197547 i \\
 0.129289\, +0.0237165 i & 0.225649\, +0.0334031 i & 0.220375 & 0.127449\, -0.00915289 i \\
 -0.0360439+0.0166259 i & 0.15571\, +0.0197547 i & 0.127449\, +0.00915289 i & 0.249643.
 \end{pmatrix}\label{matrixgad1}
\end{equation}
\end{widetext}

The ideas presented in this work extend beyond this specific implementation and can be applied to other photonic platforms, such as photonic chips\cite{keil2011all,luo2019nonlinear,fang2015nanoplasmonic}, which contain the necessary elements for constructing a fully programmable interferometer. The implementation in waveguides with electro-optical control provides a suitable platform for this proposal, offering an optical hardware capable of performing various quantum simulations.
\section{Conclusions}
We have developed a programmable interferometric platform that encodes quantum channels via photon path degrees of freedom, with polarization serving as a control parameter. By manipulating polarization, we achieve reconfigurable control over path-encoded qubits, enabling the simulation of diverse quantum channels—including phase-damping, amplitude-damping, and bit-flip channels—through adjustable optical elements. Experimental validation of the generalized amplitude damping (GAD) channel yielded a reconstructed density matrix with $95\%$ fidelity, confirming the robustness of our all-optical approach. This work establishes photonics as a versatile testbed for quantum channel simulation, offering a scalable framework to explore decoherence dynamics and fault-tolerant protocols.

While the channels discussed here model physically motivated noise processes, our platform also supports the simulation of universal Pauli channels, which combine bit-flip, phase-flip, and depolarizing noise—key for quantum error correction protocols. We provide a detailed implementation of this versatile noise model in Appendix A.

\acknowledgments
The authors acknowledge financial support from the National Council for Scientific and Technological Development (CNPq)  Grant No. 403493/2022-6 and the John Templeton Foundation Grant No. 62424. We thank Gabriel H. Aguilar and Gabriela B. Lemos for helpful discussions.

\appendix
\label{appendix}
\begin{widetext}
\section{Pauli Channel}

The \textit{Pauli channel} channel generalizes quantum noise by combining the Pauli operations \( \sigma_x \) (spin flip), \( \sigma_y \) (spin flip with phase shift), and \( \sigma_z \) (phase flip) \cite{chuang1997}. Its Kraus representation is given by
\begin{eqnarray}
\Lambda_p[\rho] = (1 - p) \rho + p \sum_{i=1}^3 q_i \sigma_i \rho \sigma_i,
\end{eqnarray}
where \( \rho \) represents the qubit state, \( p \) determines the error probability applied to the qubit state, \( q_i \) are the weights associated with the Pauli operations (with the constraint \( q_1 + q_2 + q_3 = 1 \)), and \( \sigma_i \) denotes the Pauli matrices.

In the isotropic case (\( q_1 = q_2 = q_3 = 1/3 \)), the channel reduces to the \textit{depolarizing channel}, which uniformly mixes the quantum state in all directions of the Bloch sphere.

The Pauli channel can be mapped to the interaction of a two-level system with a four-level reservoir, where the initial state is given by
\begin{eqnarray}
\rho_{SE} = \ket{\psi}\bra{\psi}\otimes\ket{00}\bra{00},
\end{eqnarray}
where \( \ket{\psi} = \alpha\ket{0}+\beta\ket{1} \) represents the system, and \( \ket{00} \) denotes the reservoir at zero temperature.
Figure \ref{Fig:estado_inicial_pauli} shows the optical circuit that implements this initial state, where the probability amplitudes are defined as \(\alpha = \cos\theta\) and \(\beta = \cos\theta\). This circuit extends the system's dimensionality compared to previous approaches, enabling the execution of the previously described operations and the encoding of states in systems with a dimension greater than four.

The map describing the evolution of these modes is shown in given by
\begin{eqnarray}
|000\rangle &\longrightarrow& \sqrt{1-p}|000\rangle + \sqrt{p} \left( \sqrt{q_1}|101\rangle + i\sqrt{q_2}|110\rangle + \sqrt{q_3}|011\rangle \right)\nonumber\\
|100\rangle &\longrightarrow& \sqrt{1-p}|100\rangle + \sqrt{p} \left( \sqrt{q_1}|001\rangle - i\sqrt{q_2}|010\rangle-\sqrt{q_3}|111\rangle \right), \label{mapa1}
\end{eqnarray}
where each mode has a probability of undergoing a transition, resulting in a mixture between the modes.

Figure \ref{Fig:canal_pauli} presents the optical circuit that implements this map, where the mixing probabilities of each mode are defined by the combination of unitary operations 1 to 6, as represented in the circuit. The parameters are given by \( p = \cos^2(\theta_1) \), \( q_1 = \cos^2(\theta_2) \), \( q_2 = \sin^2(\theta_2)\cos^2(\theta_3) \), and \( q_3 = \sin^2(\theta_2)\sin^2(\theta_3) \).

At the end of the circuit, the global state takes the form
\begin{eqnarray}
     \ket{\psi^{'}} &=& (\alpha \sqrt{1-p}|000\rangle + \beta \sqrt{1-p}|100\rangle \nonumber\\
   &+& \alpha \sqrt{p} (\sqrt{q_1}|101\rangle + i\sqrt{q_2}|110\rangle + \sqrt{q_3}|011\rangle)\nonumber\\
   &+& \beta \sqrt{p} (\sqrt{q_1}|001\rangle - i\sqrt{q_2}|010\rangle - \sqrt{q_3}|111\rangle))\ket{H}
\end{eqnarray}
where, after tracing out the reservoir degrees of freedom and polarization, the resulting system state is obtained the Eq.(\ref{matrixgad})

\begin{equation}
\rho_{s}^{'} \equiv
 \begin{pmatrix}
(1-p)|\alpha|^2 + p(q_1|\beta|^2 + q_2|\beta|^2 + q_3|\alpha|^2) & (1-p)\alpha\beta^* + p(q_1\alpha\beta^* - q_2\alpha^*\beta - q_3\alpha\beta) \\
(1-p)\alpha^*\beta + p(q_1\alpha^*\beta - q_2\alpha\beta^* - q_3\alpha^*\beta) & (1-p)|\beta|^2 + p(q_1|\alpha|^2 + q_2|\alpha|^2 + q_3|\beta|^2)
\end{pmatrix}.\label{matrixgad}
\end{equation}
The matrix above represents the quantum state \(\rho_s'\) after the application of a Pauli channel, parameterized by the probability \( p \) and the coefficients \( q_1, q_2, q_3 \), which determine the action of the Pauli operators on the initial state. This model describes the evolution of a qubit subjected to probabilistic noise, where each Pauli operator is applied with a certain weighting, leading to the transformation of the system's coherence and population components.

\begin{figure}
\centering 
\includegraphics[width=0.5\linewidth]{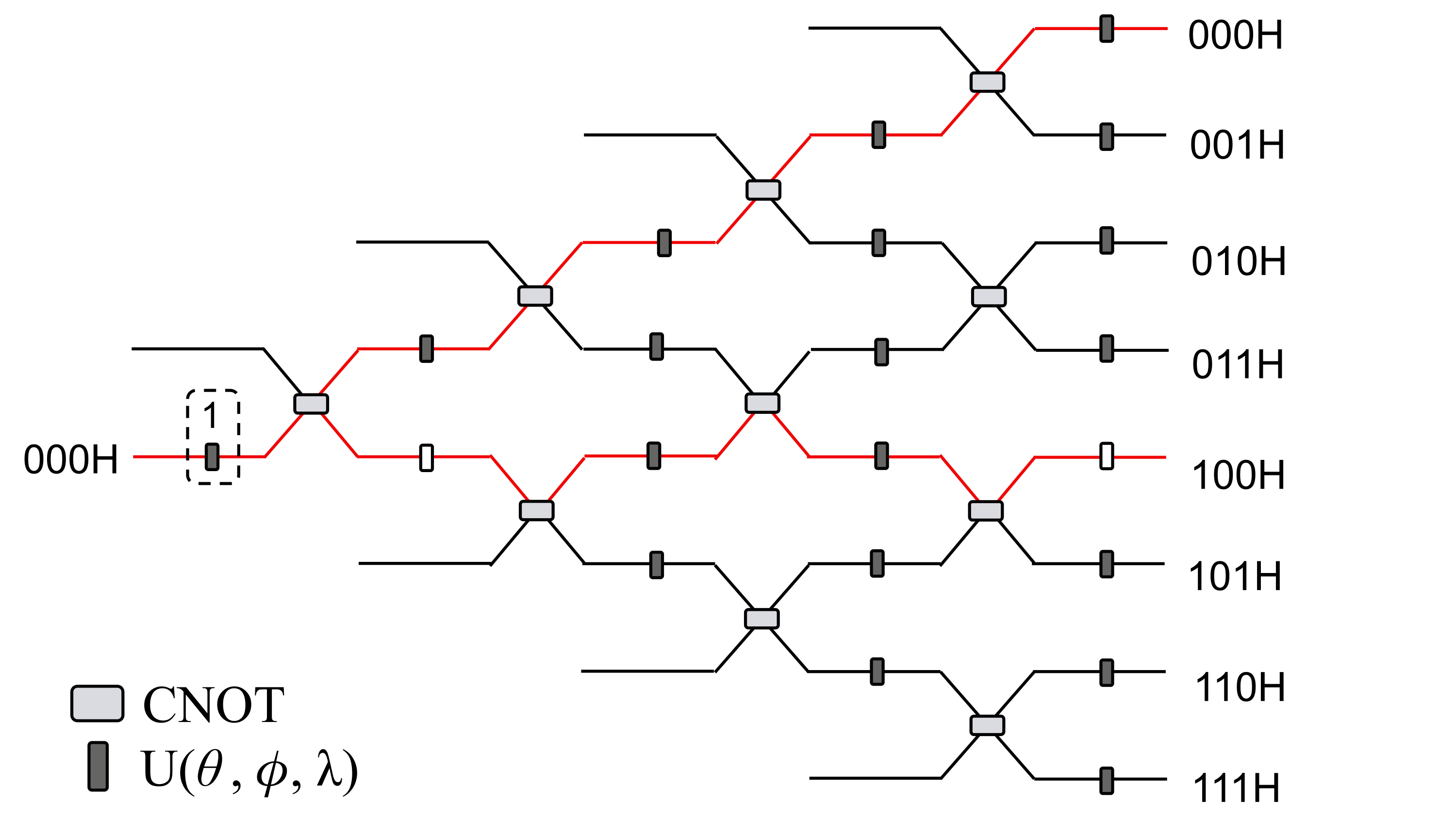}
\caption{Representative schematic of the optical circuit that implements the initial state in an 8-dimensional system. The unitary operation \( U(\theta, 0, \pi) \) controls the probability amplitude of the system's initial state, while the other unitary operations are configured as \( U(0,0,\pi) \) (represented in black) and \( U(\pi,0,\pi) \) (represented in white).}
\label{Fig:estado_inicial_pauli}
\end{figure}
\begin{figure*}[htb]
\centering 
\includegraphics[width=0.75\linewidth]{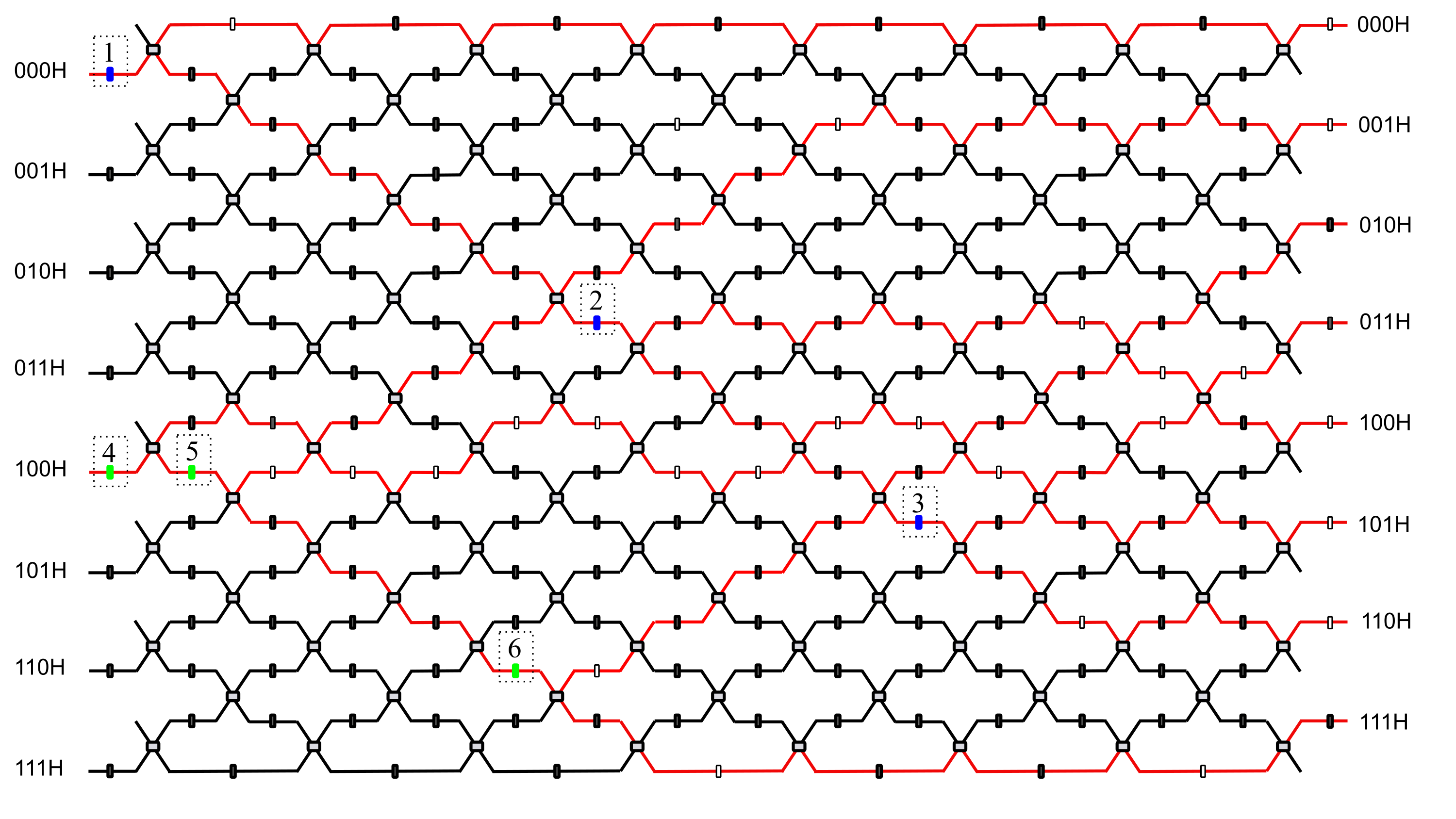}
\caption{Schematic illustration of an optical circuit featuring an arrangement of CNOT gates and local unitary gates \( U(\theta, \phi, \lambda) \). The unitary operations numbered from 1 to 6 include a free parameter \( \theta \), which controls the probabilities associated with the transitions shown in the map, while \( \phi \) and \( \lambda \) generally remain constant. The unitary gates have the following configuration: \( U_1(\theta_1, 0, \pi) \), \( U_2(\theta_2, 0, \pi) \), \( U_3(\theta_3, 0, \pi) \), \( U_4(\theta_1, 0, \pi) \), \( U_5(\theta_2, 0, \pi) \), and \( U_6(\theta_3, 0, \pi) \). }
\label{Fig:canal_pauli}
\end{figure*}

\end{widetext}


\end{document}